\documentclass[sigconf]{acmart}
\AtBeginDocument{%
  }

\usepackage{hyperref}
\usepackage{tabularx}
\usepackage{amsmath}
\usepackage{algorithm}
\usepackage{algpseudocode}
\usepackage{graphicx}
\usepackage{multirow}
\usepackage{balance}
\usepackage[table]{xcolor}
\usepackage{booktabs}
\usepackage{caption}
\usepackage{enumitem}
\copyrightyear{2026}
\acmYear{2026}
\setcopyright{cc}
\setcctype{by}
\acmConference[ICCAD '26]{IEEE/ACM International Conference on Computer-Aided Design}{November 08--12, 2026}{San Jose, CA, USA}
\acmBooktitle{IEEE/ACM International Conference on Computer-Aided Design (ICCAD '26), November 08--12, 2026, San Jose, CA, USA}
\acmDOI{10.1145/3831252.3834141}
\acmISBN{979-8-4007-2873-0/2026/11}

\begin{document}

\title[Locus: A Framework for Exploring and Optimizing Point Addition Hardware for Zero‑Knowledge Proofs]{Locus: A Framework for Exploring and Optimizing \\
Point Addition Hardware for Zero‑Knowledge Proofs}

\author{Gaurav Kuwar, Alhad Daftardar, Jianqiao Mo, Siddharth Garg, and Brandon Reagen}
\affiliation{
    \institution{New York University Tandon School of Engineering}
    \city{Brooklyn}
    \state{NY}
    \country{USA}
}
\email{\{gk2657, ajd9396, jm8782, sg175, bjr5\}@nyu.edu}


\begin{abstract}
Zero-Knowledge Proofs (ZKPs) are critical for privacy-preserving and verifiable computation, but their cryptographic primitives impose high computational overheads. One such primitive is point addition (PADD) on elliptic curves. Several prior works have implemented PADDs in hardware, but only for a few specific elliptic curves and design points, leaving a large design space unexplored, and lacking systematic guidance on hardware design trade-offs.
To address this gap, we present \emph{Locus}, a framework dedicated to optimizing and exploring point addition hardware. Given the parameters of any elliptic curve in a supported equation form, Locus automatically generates ASIC and FPGA implementations of PADD, enabling systematic exploration of the PADD design space. Using Locus, we conduct the first comprehensive hardware‑focused study of PADD designs, exploring trade‑offs over 1,000 design points.
On a 12nm technology node, our framework produces PADD designs that yield a $2.71\times$ geomean speedup and $3.11\times$ geomean area reduction compared to prior ASICs, $34.67\times$ geomean speedup over CPU, and $3.15\times$ geomean speedup on end-to-end proof generation when integrated into a prior ZKP accelerator at iso-area. Locus is available at https://github.com/cryptolets/cryptolets/tree/locus.
\end{abstract}


%

\maketitle

\begin{figure}[t]
    \centering
    \includegraphics[width=1\columnwidth]{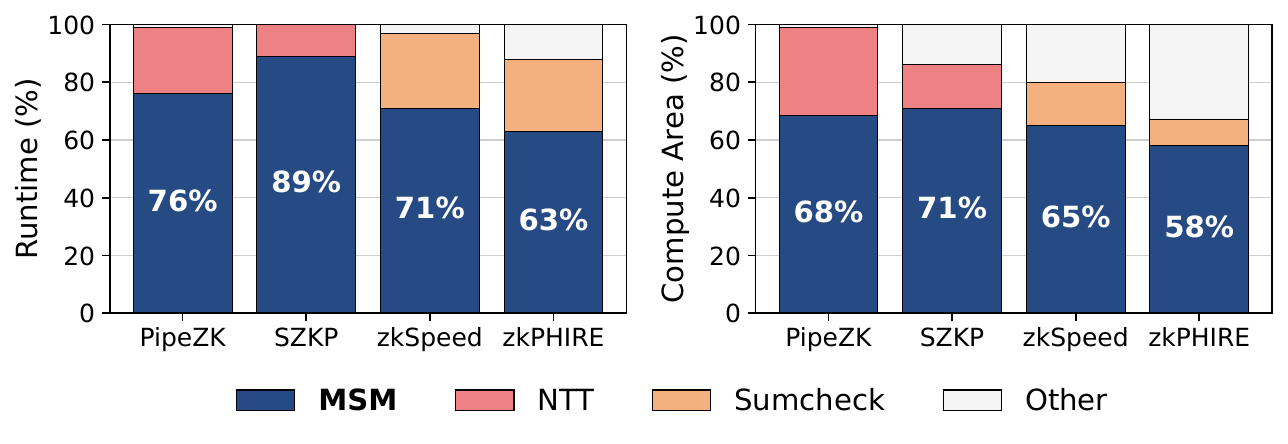}
    \vspace{-7mm}
    \caption{MSM Runtime and Compute Area breakdown in prior end-to-end ZKP accelerators at iso-problem-size ($~2^{20}$). MSM clearly dominates in both area and runtime.}
    \vspace{-2mm}
    \label{fig:zkp_breakdown}
\end{figure}

\section{Introduction}
\label{sec:intro}

\begin{figure}[t]
    \centering
    \includegraphics[width=0.9\columnwidth]{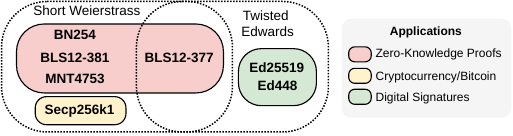}
    \caption{Curves grouped by equation form and applications.}
    \label{fig:curves}
    \vspace{-15pt}
\end{figure}

Zero-Knowledge Proofs \cite{zkp} are cryptographic protocols that enable a prover to convince a verifier that a computation was carried out correctly, without revealing any information about the private inputs. 
ZKPs have emerged as a key primitive for applications like confidential payments, rollups and validity proofs in blockchains, verifiable outsourced computation, and privacy-preserving machine learning \cite{garimella2025network}. 
As these deployments scale, prover-side cost (i.e., latency, energy, and hardware resources) quickly becomes a bottleneck, motivating dedicated hardware acceleration.

A core operation in many ZKP constructions \cite{groth, hyperplonk, orion} is the {commitment scheme}. This allows a prover to commit to a polynomial (which encodes the computation being proven) and later open it at points chosen by the verifier.
Pairing-based SNARKs such as Groth16 \cite{groth}, as well as popular commitment schemes such as KZG \cite{kzg_pcs} (used by HyperPlonk \cite{hyperplonk}) and Hyrax \cite{hyrax}, require evaluating linear combinations of elliptic curve points, implemented as multi-scalar multiplications (MSM).
Prior system-level profiling shows that MSM consistently dominates both proving time (60--90\% of runtime) and compute area (more than half) in ZKP hardware accelerators \cite{pipezk, szkp, zkspeed, zkphire, mo2025mtu}, as shown in \autoref{fig:zkp_breakdown}.
The MSM itself is built from elliptic-curve point additions (PADD). PADDs are expensive, each typically requiring tens of field multiplications \cite{pipezk, szkp, zkspeed, priormsm, cyclonemsm, zkphire}. In practical ZKP workloads, MSM sizes routinely exceed $2^{20}$ points or more, meaning the prover performs millions of PADD operations. In zkSpeed, for example, MSMs incur 4 \textit{billion} field multiplications \cite{zkspeed}. 
Consequently, ZKP provers spend the majority of their time, energy, and area inside these repeated PADD operations. 
Therefore, understanding and optimizing PADD performance is critical for scaling future ZKP systems.

As shown in \autoref{fig:curves}, elliptic-curve point addition (PADD) is a fundamental building block not only in ZKPs but also in blockchain signature verification (e.g., BLS, ECDSA), secure messaging, and key-exchange protocols. Despite this broad relevance, prior hardware accelerators typically develop one highly specialized PADD design per curve. These implementations provide useful curve-specific optimizations for ASIC \cite{szkp, pipezk, zkspeed, zkphire, priormsm} and FPGA \cite{cyclonemsm, hardcamlmsm, bstmsm, msmac} but offer no unified methodology for exploring microarchitectural trade-offs across curves, design choices, or hardware targets.

This fragmentation is increasingly limiting. While cryptographic algorithms are often compared using asymptotic complexity, real hardware behavior depends on additional factors (including pipeline depth, area, and on-chip memory) that asymptotic models do not capture. As a result, theoretical cost alone cannot predict the actual performance of PADD in modern architectures, and designers lack a hardware-grounded way to reason about curve- and microarchitecture-level choices.

To address this gap, we introduce \emph{Locus}, a framework for exploring and optimizing point addition and MSM hardware for ZKPs and other elliptic-curve applications. 
Locus implements highly optimized primitives to construct efficient PADD units, systematically characterizes the PADD design space across many curves and microarchitectural choices, and generates synthesizable RTL for ASIC and FPGA backends. Locus makes the following contributions:

\begin{itemize}[leftmargin=*]
    \item 
    We perform the first systematic hardware-centric study of elliptic-curve point addition across a broad set of select curves used in ZKPs, blockchain validation, and digital signatures. 
    Locus explores over 1,000 design points by sweeping key PADD-specific design optimizations such as multiplier decomposition, systematic fixing of constants, etc.
    \item We provide an extensible toolchain that takes curve parameters and design configurations as input, automatically generating RTL with comparable ASIC/FPGA metrics and enabling rapid exploration of curve- and microarchitecture-level choices.
    \item We integrate Locus PADDs into full MSM pipelines and a prior end-to-end ZKP accelerator to evaluate system-level behavior under large ZKP workloads, showing how PADD-level optimizations translate to improved MSM and proving performance.
    \item We generate PADD designs with 2.71$\times$ geomean speedup and 3.11$\times$ area reduction over prior ASICs, $34.67\times$ geomean speedup over CPU, $2.72\times$ speedup over prior ASIC MSMs at iso-area, and $3.15\times$ geomean speedup on end-to-end proof generation when integrated into a prior ZKP accelerator at iso-area.
\end{itemize}

\section{Background}
\subsection{Elliptic Curves}
Elliptic curves are sets of points $(x, y)$ over a finite field $\mathbb{F}_q$ (where $q$ is a large prime) defined by a non-singular algebraic equation. 
Two common \textit{equation forms} in cryptography are 
Short Weierstrass: 
$y^2 = x^3 + ax + b$,
and Twisted Edwards: 
$ax^2 + y^2 = 1 + dx^2y^2$.
Our framework provides plug-and-play support for both forms and is easily extensible to others. Cryptographers have established standardized \textit{named curves}, each with a purposefully chosen prime field modulus $q$ and curve coefficients, whose security has been extensively studied. Therefore, although Locus supports \textit{any} curve in these forms (and is extensible to other forms), in this paper, we focus our evaluation on a set of these standardized and relevant named curves spanning both forms, as shown in \autoref{fig:curves}.

\vspace{-2mm}
\subsection{Multi-Scalar Multiplication}
MSMs are dot products between a vector of scalars and a vector of 2D/3D points on an elliptic curve.
MSMs are expensive because point multiplication is achieved by repeated point addition, and the scalars are typically 256 - 753 bits.
\textit{Pippenger's Algorithm} \cite{pippenger} is often used to  reduce the computational cost by partitioning scalar multiplications across parallel \textit{windows}, or chunks of wide scalars, followed by combining reduction steps. Several prior works use this technique \cite{szkp, zkphire, zkspeed, priormsm, pipezk, bstmsm, hardcamlmsm, cyclonemsm, gypsophila, legozk, myosotis, priormsm, distmsm, morph}. 
However, elliptic-curve point coordinates are \textit{also} 256 - 753 bits, requiring expensive large-bitwidth modular arithmetic for a single point addition.

\vspace{-2mm}
\subsection{Point Addition}
Point addition on an elliptic curve is geometrically defined as follows: given two distinct points $P(x, y)$ and $Q(x, y)$, draw a line through them; it intersects the curve at a third point $-R$, and $R = P + Q$ is the reflection of $-R$ over the $x$-axis. In the case where $P = Q$, we must perform a different operation called point doubling (PDBL), which uses the tangent line at $P$ instead.

The standard point addition formula, using affine coordinates $(x, y)$, requires a computationally expensive modular inversion. To avoid this cost, points can be transformed into alternative coordinate systems that trade the inversion for additional, but relatively cheaper, modular multiplications. This transformation modifies the algebraic formulas for addition and doubling. We implement Short Weierstrass curves in Jacobian coordinates $(X, Y, Z)$ and Twisted Edwards curves in Extended Projective coordinates $(X, Y, Z, T)$, each requiring its own formula. For these formulas, we reference the Explicit-Formulas Database \cite{hyperelliptic}, which catalogs efficient point addition formulas across coordinate systems.

Most ZKP protocols use Short Weierstrass curves, whose point addition formulas distinguish addition and doubling, necessitating runtime branching and an equality check. On the other hand, Twisted Edwards curves tend to use unified addition formulas, which use less modular multiplications. Prior works \cite{priormsm, cyclonemsm, hardcamlmsm, bstmsm, myosotis} leverage \textit{birational equivalence}, a mapping that allows BLS12-377 (a popular Short Weierstrass curve for ZKPs) to be expressed in the more efficient Twisted Edwards form. We denote this variant as BLS12-377* throughout the paper.

\vspace{-2mm}
\subsection{Related Work}
Prior works on MSM acceleration optimize PADD units using curve-specific multiplier configurations. Several target FPGAs; MSMAC \cite{msmac} applies Karatsuba multipliers for the BN254 (or BN128) curve on FPGA, while CycloneMSM, HardcamlMSM, and BSTMSM implement BLS12-377* PADDs using Karatsuba and constant multipliers \cite{cyclonemsm, bstmsm, hardcamlmsm}.
Others use ASIC techniques. PriorMSM \cite{priormsm} uses Karatsuba-based PADDs for the BLS12-377* curve, while SZKP \cite{szkp} and zkSpeed \cite{zkspeed} use high-level synthesis tools to synthesize curve-specific PADDs for BN254, MNT4753, and BLS12-381. PipeZK \cite{pipezk} targets the same curves with custom, hand-tuned PADDs.

Each of these works consequently builds MSM accelerators around a particular hand-tuned PADD configuration, leaving limited exploration of alternative PADD microarchitectures that may offer different trade-offs in latency, area, and throughput. As a result, the broader PADD design space—and its impact on MSM efficiency—has not been systematically characterized.

\section{The Locus Framework}
Locus is an extensible, plug-and-play framework for generating optimized PADD designs across elliptic-curve equation forms and \textit{any} curve parameters. 
Users can simply provide the sweep specification and curve parameters in configuration files.
Locus automates the entire flow, from design generation to verification and performance analysis, while evaluating multiple designs in parallel across available CPU threads to enable large-scale design-space exploration. The framework's modular structure allows easy extension to other equation forms and future development.

Locus is built on top of High-Level Synthesis (HLS), specifically Catapult HLS, which converts high-level languages into Register-Transfer Level (RTL)  hardware description languages. Prior ZKP accelerators \cite{szkp, zkspeed, zkphire} also use HLS, but largely treat it as a black box: they rely on default flows and built-in primitives (e.g., Catapult's native multipliers) to obtain rough area estimates, leaving significant performance on the table. In contrast, Locus implements its own optimized cryptographic primitives, including large-bitwidth multipliers, modular arithmetic, and point addition units, as parameterized HLS source code. This allows us to expose core microarchitectural choices as explicit design knobs and apply targeted HLS-level optimizations, producing designs that are significantly more efficient than prior works.

Point addition is a directed acyclic graph (DAG) of modular multiplications, additions, and subtractions. Among these, modular multiplication (Modmul) dominates the area and latency of the PADD unit, as well as other cryptographic workloads \cite{rpu, osiris, ciflow, nocap}. Locus targets the key bottlenecks within Modmul, enabling it to compose highly optimized PADDs. 

\subsection{Addressing the Multiplier Bottleneck}\label{sec:mul}
A direct Modmul ($(A \times B) \bmod q$) requires an expensive division to reduce modulo $q$. Montgomery \cite{montgomery} and Barrett \cite{barrett} reduction are two standard algorithms that eliminate this division. Locus implements both as separate kernels. Montgomery reduction uses a precomputed constant $q'$ and requires operands to be in the Montgomery domain, while Barrett requires only a precomputed constant $\mu$, but using a multiplier with $2\times$ the Modmul operand bitwidth. Each exposes different area and latency trade-offs depending on the modulus $q$, so Locus exposes the choice as a design knob.

Even with an optimized reduction algorithm, each Modmul still requires three large-bitwidth integer multiplications, which remain the critical hardware bottleneck.

\subsubsection{Decomposed Multipliers.} 
Karatsuba \cite{karatsuba1995complexity} and Schoolbook are techniques that decompose large multiplications into smaller partial products using algebraic identities. Schoolbook splits each operand into two halves, producing \textit{four} partial products. Karatsuba reduces this to \textit{three}, trading one multiplication for two additions and two subtractions. These decompositions can be applied recursively, breaking partial products down further at each level. While Karatsuba achieves lower area due to fewer partial products, the additional additions and subtractions typically result in higher latency, as they lie on the critical path. Moreover, decomposition cannot be applied indefinitely; each level adds recombination overhead (additions, shifts, adder trees), and beyond a certain recursive depth, area can actually increase. At the same time, Schoolbook is not always the fastest choice: depending on the bitwidth and timing constraints, Karatsuba at a specific depth can match Schoolbook's latency while using less area. The optimal depth and decomposition strategy therefore depends on operand bitwidth. Since different curves operate at different bitwidths, a fixed choice could be suboptimal across curves, motivating Locus to expose these as design knobs. While prior works \cite{cyclonemsm, hardcamlmsm, bstmsm, msmac, priormsm, gypsophila} use these techniques, they rely on static, hand-picked configurations.

\begin{figure}[t]
\centering
\includegraphics[width=1.0\columnwidth]{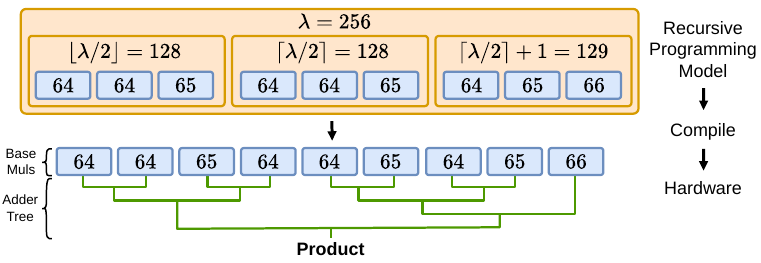}
\vspace{-20pt}
\caption{Decomposed Karatsuba Multiplier for operand width $\lambda = 256$ bits at Karatsuba Depth = 2 and Base Width = 64 bits. We include details on how deeper levels' multiplier bitwidths are derived mathematically.}
\label{fig:kar}
\vspace{-15pt}
\end{figure}

To address these challenges, Locus implements custom parameterized multipliers using a 3-layer recursive model, where each layer uses a different decomposition strategy: Karatsuba $\rightarrow$ Schoolbook $\rightarrow$ Baseline (Catapult's native multiplier). Starting from the top level, Locus applies Karatsuba decomposition for the configured number of recursion levels (\textit{Karatsuba Depth}), then switches to Schoolbook, and finally to Baseline multipliers once operands reach a configurable \textit{Base Width} ($\lessapprox$ 64 bits for ASICs), where Catapult's native multipliers are generally efficient. A \textit{Karatsuba Depth} of 0 corresponds to pure Schoolbook decomposition, with higher depth progressively reducing the number of partial products. Locus exposes both \textit{Karatsuba Depth} and \textit{Base Width} as explicit design knobs. \autoref{fig:kar} shows how this model is compiled into a hybrid Karatsuba multiplier, with baseline multipliers in parallel followed by a pipelined adder tree at the hardware level. 

Squaring ($A \times A$) similarly reduces partial products without the additional overhead of Karatsuba, by exploiting the symmetry of equal operands. Locus adds a squaring layer on top of the existing 3-layer multiplier model, generating optimized squaring primitives that compose into Modular Squaring (Modsq) units. This is particularly valuable since approximately one-third of operations in Short Weierstrass PADD formulas are Modsqs.

\subsubsection{Constant Multipliers}
Two out of the three multiplications in a Modmul involve constant parameters. When these constants are known at design time, the corresponding variable multiplier can be replaced with a constant multiplier, which uses the shift-add method, where multiplications by 0-bits in the constant are compiled away, hence lower Hamming weight constants produce more efficient hardware. \textit{Non-Adjacent Form} (NAF) encoding, a signed-bit representation that minimizes the number of non-zero digits, further reduces the Hamming weight. For low latency, these constant multipliers are implemented as adder trees. Locus leverages Catapult's native support for constant multiplier generation. We denote replacing a variable multiplier with a constant multiplier as \textit{fixing} that parameter, and \textit{variable} otherwise. Whether to fix a constant is not a straightforward choice: depending on the parameter's Hamming weight, the area and latency of the constant multiplier can vary. This makes the optimal fixing strategy curve-dependent, motivating Locus to expose it as a design knob. Additionally, PADD formulas sometimes involve multiplication by curve coefficients ($a$, $d$, $k=2d$), which can additionally be fixed. In total, Locus exposes the reduction constants ($q'$, $\mu$), the prime modulus ($q$), and the curve coefficients ($a$, $d$, $k$) each independently as \textit{fixed} or \textit{variable}.

\subsubsection{HLS Optimizations}
In decomposed multipliers, adder trees dominate latency, and constant multipliers rely entirely on adder trees. Catapult HLS has a  \textit{Clustering} feature that replaces regular adders with Carry Save Adders (CSAs) with much lower, and constant latencies across bitwidths at only moderate area costs. We find that PADDs with CSAs achieve up to $1.5\times$ speedup. 

\vspace{-1mm}
\subsection{Multi-Precision} \textit{Multi-precision} arithmetic is an alternative approach where one or more smaller, shared \textit{word width} arithmetic units are used to compute larger bitwidth arithmetic, as opposed to full-bitwidth \textit{single-precision}. This provides large area savings, but with a heavy latency and throughput penalty. CPUs \cite{gmp, zhang_cpu} and GPUs \cite{zhang_gpu} use multi-precision for 
cryptographic arithmetic. 
In Locus, we implement textbook \cite{mp_src} multi-precision modular primitives for comparison.

\vspace{-1mm}
\subsection{The Point Addition Unit}
With Locus's optimized modular arithmetic primitives in place, Locus composes them into complete PADD and PDBL units, where all design knobs come together. Each PADD formula is implemented using these primitives, supporting both single-precision and multi-precision designs. For each curve and configuration, Locus generates a PADD design with area and latency metrics, enabling design-space exploration across curves to identify optimal configurations.

\begin{algorithm}[b]
\caption{MSM Latency Model}
\label{alg:msm}
\begin{algorithmic}[1]
\State $T_{\text{sort}} = P \cdot \frac{\lambda / w}{K}$ \Comment{Bucket Sorting}
\vspace{1mm}
\State $T_{\text{br}} = (2t_{\text{add}} \cdot (2^{w} - 1)) \cdot \frac{\lambda/w}{K}$ \Comment{Bucket Reduction}
\vspace{1mm}
\State $T_{\text{wr}} = (K \cdot w \cdot t_{\text{dbl}}) \cdot \frac{\lambda / w}{K}$ \Comment{Window Reduction}
\vspace{1mm}
\State $T = T_{\text{sort}} + T_{\text{br}} + T_{\text{wr}}$ \Comment{Total}
\vspace{1mm}
\State Find $w^\ast = \arg\min_{w}\; T(w, P, \lambda, K, t_{add}, t_{dbl})$ 
\end{algorithmic}
\end{algorithm}

\vspace{-1mm}
\subsection{Optimizing MSM Performance}\label{sec:locus_msm}

With optimized PADD and PDBL units in hand---the core building blocks for MSM acceleration---Locus evaluates their impact on end-to-end MSM performance. We first identify Pareto-optimal PADDs from the design space and use their area and latency metrics in a lightweight analytical MSM model. Algorithm~\ref{alg:msm} presents a simplified, first-order latency model for dense Pippenger MSMs following SZKP's architecture \cite{szkp}. The model assumes uniform scalars and an SZKP-style scheduler that serially accumulates bucket partial sums; it does not explicitly model more aggressive priority-based or greedy scheduling strategies \cite{pipezk, priormsm, gypsophila}. The algorithm parameters are the scalar bitwidth $\lambda$ (set by the curve), the window size $w$, and the number of points $P$. The hardware parameters are the number of parallel MSM PEs $K$ and the pipeline depths $t_{\text{add}}$ and $t_{\text{dbl}}$ of the PADD and PDBL units. Setting all three to $1$ estimates the number of PADD operations required by the MSM. Lastly, while Algorithm~\autoref{alg:msm} shows only the compute-side formulation, our evaluation accounts for memory bandwidth; prior work reports 30--80~GB/s bandwidth demand and finds these MSMs remain compute-bound under DDR/HBM constraints \cite{szkp, zkspeed, zkphire}.

In MSM architectures, the optimal window size $w^\ast$ is often chosen to minimize PADD operations. While sorting dominates, $w$ cannot be made arbitrarily large: increasing $w$ reduces the number of windows but exponentially increases bucket-reduction work. The optimal $w^\ast$ shifts when latency costs (e.g., $t_{\text{add}}, t_{\text{dbl}}$) and area constraints are included, motivating a hardware-aware search.

MSM accelerators typically select the window size after constructing the MSM architecture and running time-consuming, cycle-accurate simulations over workloads of length $2^{20}$--$2^{24}$ points \cite{bstmsm, cyclonemsm, hardcamlmsm, szkp, zkspeed, gypsophila, priormsm}. Because these evaluations already sweep a large parameter space (e.g., window size and number of PEs), prior works generally evaluate a single PADD design, often specialized to a particular curve, leaving PADD microarchitectural variation largely unexplored. In contrast, our analytical model prunes over $10{,}000$ MSM configurations in under $10$ seconds and remains within $5\%$ of cycle-accurate simulation for the evaluated SZKP-style configurations. Pareto-optimal candidates are then validated with cycle-accurate simulation and synthesis, enabling a substantially larger joint algorithm--hardware search.

\begin{table}[t]
\centering
\caption{\small PADD Design Space for ASIC}
\label{tab:design_space_padd}
\vspace{-3mm}
\small
\resizebox{\columnwidth}{!}{
\setlength{\tabcolsep}{1mm}{
\begin{tabular}{|c|c|}
\hline
\textbf{Design Setting} & \textbf{Values} \\ \hline \hline
Curve Type & BN254, BLS12-377, BLS12-381, MNT4753, \\
& Secp256k1, Ed25519, Ed448, BLS12-377* \\ \hline
Modmul Algorithm & Montgomery, Barrett \\ \hline
Multiplier Type & Baseline, Schoolbook, Karatsuba \\ \hline
Karatsuba Depth & 1, 2, 3, 4 \\ \hline
Prime ($q$) Type & Fixed, Variable \\ \hline
Reduction Const Type & Fixed, Variable \\ \hline
Equation Coefficient Type & Fixed, Variable \\ \hline
Precision Type & Single-Precision, Multi-Precision \\ \hline
Multi-Prec word width & 16, 32, 64, 128 \\ \hline
\end{tabular}
}}
\vspace{-3mm}
\end{table}

\section{Experimental Setup}
Using Locus, we perform a comprehensive design space sweep across parameter combinations shown in \autoref{tab:design_space_padd}. Locus supports Initiation Interval
(II) $>1$ designs, but we focus our evaluation on fully-pipelined $\text{II}=1$ designs for single-precision implementations. For baseline PADDs, we use baseline (Catapult's native) multipliers and keep all constants variable, as in prior works \cite{szkp}.

\textbf{ASIC Implementation. }
We use Siemens Catapult HLS~2025.1 with a GF 12nm library for all PADD designs, targeting a 1\,GHz clock (except designs with baseline multiplier designs at $\geq 512$ bits, clocked at 667\,MHz). We use Synopsys Design Compiler 2026.03 for area and power estimates.
For MSM experiments, we use the latency model from Algorithm~\ref{alg:msm} together with on-chip memory area estimates from the GF12 SRAM compiler to perform the initial design-space sweeps. We validate selected design points using cycle-accurate simulation following the methodology of \cite{szkp, mo2023haac}, and find our model is within $5\%$ of actual latency.

\textbf{FPGA Implementation. } We use Vivado v2024.2 for RTL synthesis at varying clock speeds. All designs are evaluated on two representative AMD-Xilinx FPGA platforms: Versal HBM (VH1782) for state-of-the-art capabilities, and UltraScale+ (VU9P) for comparison with established literature benchmarks.

\textbf{Testing and Verification.} Locus validates all generated designs, including PADD units and all primitives (modular and integer arithmetic units), against software reference implementations within a custom cryptographic library. The library is implemented in Python and generates cryptographically sound random and edge-case inputs for each kernel. This verification is fully integrated into Locus's automated pipeline, which generates a configurable number of test samples with golden models and verifies both the C++ HLS source code with the Open SystemC Initiative (OSCI) flow and generated RTL with Siemens QuestaSim 2026.2.
All designs in our evaluation have been validated through this pipeline using 1,000 samples.

\begin{table}[b]
\vspace{-2mm}
\centering
\caption{Design Knobs for Best Performance Designs}
\vspace{-1em}
\label{tab:design_knobs}
\footnotesize
\resizebox{1\columnwidth}{!}{
\setlength{\tabcolsep}{1mm}{
\begin{tabular}{|c|c|c|c|c|c|c|c|c|c|c|c|c|c|c|c|c|}
\hline
\multirow{4}{*}{\textbf{Curve}} & \multicolumn{8}{c|}{\textbf{Montgomery}} & \multicolumn{8}{c|}{\textbf{Barrett}} \\ \cline{2-17}
& \multicolumn{4}{c|}{\textbf{Fastest}} & \multicolumn{4}{c|}{\textbf{Smallest}} & \multicolumn{4}{c|}{\textbf{Fastest}} & \multicolumn{4}{c|}{\textbf{Smallest}} \\ \cline{2-17}
& \rotatebox{90}{\textbf{ct}} & \rotatebox{90}{\textbf{qt}} & \rotatebox{90}{\textbf{rct}} & \rotatebox{90}{\textbf{kar}} & \rotatebox{90}{\textbf{ct}} & \rotatebox{90}{\textbf{qt}} & \rotatebox{90}{\textbf{rct}} & \rotatebox{90}{\textbf{kar}} & \rotatebox{90}{\textbf{ct}} & \rotatebox{90}{\textbf{qt}} & \rotatebox{90}{\textbf{rct}} & \rotatebox{90}{\textbf{kar}} & \rotatebox{90}{\textbf{ct}} & \rotatebox{90}{\textbf{qt}} & \rotatebox{90}{\textbf{rct}} & \rotatebox{90}{\textbf{kar}} \\ \hline
BN254 & - & \cellcolor[HTML]{c84030}\textcolor{white}{F} & \cellcolor[HTML]{FFDAB9}{V} & \cellcolor[HTML]{97d6b9}{0} & - & \cellcolor[HTML]{FFDAB9}{V} & \cellcolor[HTML]{c84030}\textcolor{white}{F} & \cellcolor[HTML]{1f80b8}\textcolor{white}{2} & - & \cellcolor[HTML]{FFDAB9}{V} & \cellcolor[HTML]{c84030}\textcolor{white}{F} & \cellcolor[HTML]{97d6b9}{0} & - & \cellcolor[HTML]{FFDAB9}{V} & \cellcolor[HTML]{c84030}\textcolor{white}{F} & \cellcolor[HTML]{1f80b8}\textcolor{white}{2} \\ \hline
BLS12-377 & - & \cellcolor[HTML]{c84030}\textcolor{white}{F} & \cellcolor[HTML]{c84030}\textcolor{white}{F} & \cellcolor[HTML]{1f80b8}\textcolor{white}{2} & - & \cellcolor[HTML]{FFDAB9}{V} & \cellcolor[HTML]{c84030}\textcolor{white}{F} & \cellcolor[HTML]{24419a}\textcolor{white}{3} & - & \cellcolor[HTML]{c84030}\textcolor{white}{F} & \cellcolor[HTML]{c84030}\textcolor{white}{F} & \cellcolor[HTML]{1f80b8}\textcolor{white}{2} & - & \cellcolor[HTML]{FFDAB9}{V} & \cellcolor[HTML]{c84030}\textcolor{white}{F} & \cellcolor[HTML]{24419a}\textcolor{white}{3} \\ \hline
BLS12-381 & - & \cellcolor[HTML]{c84030}\textcolor{white}{F} & \cellcolor[HTML]{c84030}\textcolor{white}{F} & \cellcolor[HTML]{1f80b8}\textcolor{white}{2} & - & \cellcolor[HTML]{FFDAB9}{V} & \cellcolor[HTML]{c84030}\textcolor{white}{F} & \cellcolor[HTML]{24419a}\textcolor{white}{3} & - & \cellcolor[HTML]{c84030}\textcolor{white}{F} & \cellcolor[HTML]{c84030}\textcolor{white}{F} & \cellcolor[HTML]{1f80b8}\textcolor{white}{2} & - & \cellcolor[HTML]{FFDAB9}{V} & \cellcolor[HTML]{c84030}\textcolor{white}{F} & \cellcolor[HTML]{24419a}\textcolor{white}{3} \\ \hline
MNT4753 & - & \cellcolor[HTML]{FFDAB9}{V} & \cellcolor[HTML]{c84030}\textcolor{white}{F} & \cellcolor[HTML]{97d6b9}{0} & - & \cellcolor[HTML]{FFDAB9}{V} & \cellcolor[HTML]{c84030}\textcolor{white}{F} & \cellcolor[HTML]{081d58}\textcolor{white}{4} & - & \cellcolor[HTML]{c84030}\textcolor{white}{F} & \cellcolor[HTML]{c84030}\textcolor{white}{F} & \cellcolor[HTML]{24419a}\textcolor{white}{3} & - & \cellcolor[HTML]{FFDAB9}{V} & \cellcolor[HTML]{FFDAB9}{V} & \cellcolor[HTML]{081d58}\textcolor{white}{4} \\ \hline
Secp256k1 & - & \cellcolor[HTML]{c84030}\textcolor{white}{F} & \cellcolor[HTML]{FFDAB9}{V} & \cellcolor[HTML]{97d6b9}{0} & - & \cellcolor[HTML]{c84030}\textcolor{white}{F} & \cellcolor[HTML]{c84030}\textcolor{white}{F} & \cellcolor[HTML]{1f80b8}\textcolor{white}{2} & - & \cellcolor[HTML]{c84030}\textcolor{white}{F} & \cellcolor[HTML]{c84030}\textcolor{white}{F} & \cellcolor[HTML]{97d6b9}{0} & - & \cellcolor[HTML]{c84030}\textcolor{white}{F} & \cellcolor[HTML]{c84030}\textcolor{white}{F} & \cellcolor[HTML]{1f80b8}\textcolor{white}{2} \\ \hline
Ed25519 & \cellcolor[HTML]{c84030}\textcolor{white}{F} & \cellcolor[HTML]{c84030}\textcolor{white}{F} & \cellcolor[HTML]{c84030}\textcolor{white}{F} & \cellcolor[HTML]{40b5c4}\textcolor{white}{1} & \cellcolor[HTML]{c84030}\textcolor{white}{F} & \cellcolor[HTML]{c84030}\textcolor{white}{F} & \cellcolor[HTML]{c84030}\textcolor{white}{F} & \cellcolor[HTML]{1f80b8}\textcolor{white}{2} & \cellcolor[HTML]{c84030}\textcolor{white}{F} & \cellcolor[HTML]{c84030}\textcolor{white}{F} & \cellcolor[HTML]{c84030}\textcolor{white}{F} & \cellcolor[HTML]{97d6b9}{0} & \cellcolor[HTML]{FFDAB9}{V} & \cellcolor[HTML]{c84030}\textcolor{white}{F} & \cellcolor[HTML]{c84030}\textcolor{white}{F} & \cellcolor[HTML]{1f80b8}\textcolor{white}{2} \\ \hline
BLS12-377$^*$ & \cellcolor[HTML]{c84030}\textcolor{white}{F} & \cellcolor[HTML]{c84030}\textcolor{white}{F} & \cellcolor[HTML]{c84030}\textcolor{white}{F} & \cellcolor[HTML]{1f80b8}\textcolor{white}{2} & \cellcolor[HTML]{FFDAB9}{V} & \cellcolor[HTML]{FFDAB9}{V} & \cellcolor[HTML]{c84030}\textcolor{white}{F} & \cellcolor[HTML]{24419a}\textcolor{white}{3} & \cellcolor[HTML]{c84030}\textcolor{white}{F} & \cellcolor[HTML]{c84030}\textcolor{white}{F} & \cellcolor[HTML]{c84030}\textcolor{white}{F} & \cellcolor[HTML]{1f80b8}\textcolor{white}{2} & \cellcolor[HTML]{FFDAB9}{V} & \cellcolor[HTML]{FFDAB9}{V} & \cellcolor[HTML]{c84030}\textcolor{white}{F} & \cellcolor[HTML]{24419a}\textcolor{white}{3} \\ \hline
Ed448 & \cellcolor[HTML]{c84030}\textcolor{white}{F} & \cellcolor[HTML]{c84030}\textcolor{white}{F} & \cellcolor[HTML]{c84030}\textcolor{white}{F} & \cellcolor[HTML]{40b5c4}\textcolor{white}{1} & \cellcolor[HTML]{c84030}\textcolor{white}{F} & \cellcolor[HTML]{c84030}\textcolor{white}{F} & \cellcolor[HTML]{c84030}\textcolor{white}{F} & \cellcolor[HTML]{24419a}\textcolor{white}{3} & \cellcolor[HTML]{c84030}\textcolor{white}{F} & \cellcolor[HTML]{c84030}\textcolor{white}{F} & \cellcolor[HTML]{c84030}\textcolor{white}{F} & \cellcolor[HTML]{1f80b8}\textcolor{white}{2} & \cellcolor[HTML]{c84030}\textcolor{white}{F} & \cellcolor[HTML]{c84030}\textcolor{white}{F} & \cellcolor[HTML]{c84030}\textcolor{white}{F} & \cellcolor[HTML]{24419a}\textcolor{white}{3} \\ \hline
\end{tabular}
}
}
\small \textbf{ct}=curve coefficient type, \textbf{qt}=prime q type, \textbf{rct}=reduction constant type, \textbf{V}=variable, \textbf{F}=fixed, \textbf{kar}=karatsuba depth.
\end{table}

\section{Evaluation Results}
In this section, we explore the PADD design space, optimizations, and insights on ASIC designs. We quantify the impact of each optimization through an ablation study and by analyzing key design knobs in isolation. We then compare against prior ASIC, FPGA, and CPU implementations and showcase rapid exploration of optimized PADD formulas. Finally, we conduct a case study on MSM, and evaluate on an end-to-end ZKP accelerator using Locus PADDs.

\subsection{Pareto Space Analysis}
\begin{figure}[t]
\centering
\includegraphics[width=\columnwidth]{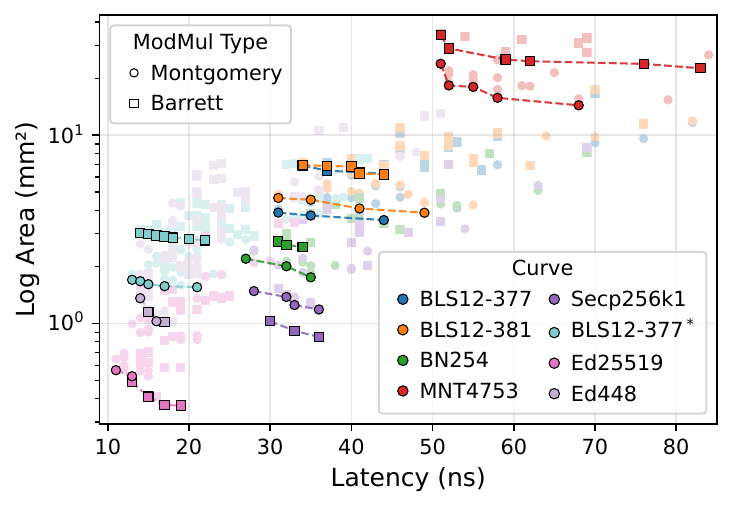}
\vspace{-22pt}
\caption{Design Space of PADD across named curves with Pareto-optimal points highlighted.}
\label{fig:design_space_padd}
\vspace{-5mm}
\end{figure}

\autoref{fig:design_space_padd} shows the large design space for all fully-pipelined, single-precision PADDs across named curves, highlighting Locus's ability to explore this space and identify Pareto-optimal designs. We analyze Montgomery and Barrett Modmul types separately, finding Pareto-optimal points for each.

Montgomery-based designs \textit{always} yield the fastest implementation across all evaluated curves and also produce smaller designs for the majority of cases (5 of 8). However, for a select few curves (Ed25519 and Secp256k1), Barrett achieves roughly $30\%$ smaller designs compared to Montgomery. This is largely due to the curve's prime modulus, which yields a Barrett reduction constant ($\mu$) with a lower NAF Hamming weight than Montgomery's ($q'$), and therefore a more area-efficient constant multiplier.

When comparing the fastest and smallest Pareto-optimal designs, several curves exhibit proportional area-latency trade-offs. However, certain curves deviate from this trend. For BLS12-377 and BLS12-381 using Montgomery Modmuls, the trade-off favors the fastest design: we achieve more speedup ($42\%$ and $58\%$) compared to area increase ($10\%$ and $20\%$). Conversely, for MNT4753 using Montgomery Modmuls, the trade-off favors the smallest design.

\autoref{tab:design_knobs} shows which settings yield optimal designs. We find that Schoolbook doesn't always result in the fastest optimal design, but low to mid range Karatsuba depth does. The smallest designs consistently use the deepest Karatsuba depth available for the curve's bitwidth. The curve- and reduction- specific constants have an impact on optimal Karatsuba depth: for Ed25519, Ed448, and MNT4753, the Karatsuba depth differs for the fastest designs across Montgomery and Barrett Modmuls. For Barrett, fixed reduction constants always result in the fastest designs because Barrett reduction constants are 2$\times$ the bitwidth of the target datatype.

\begin{figure}[b]
\vspace{-4mm}
\centering\includegraphics[width=\columnwidth]{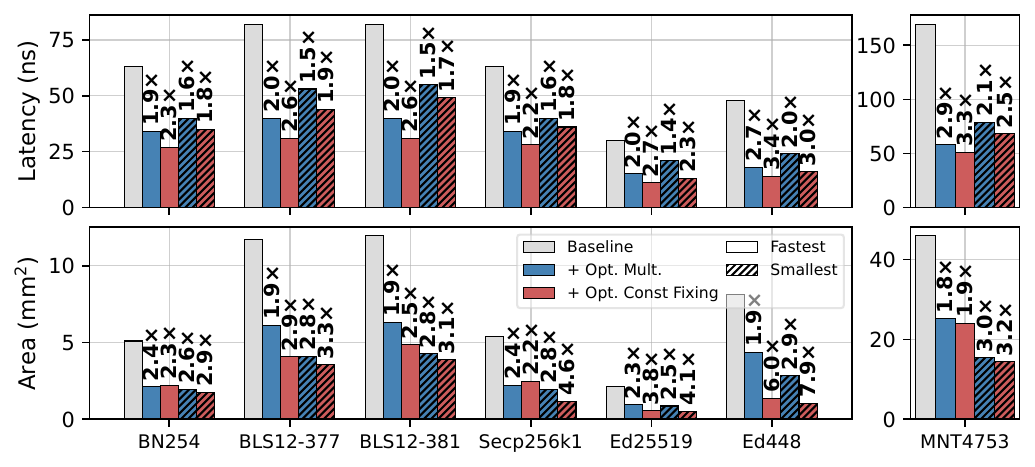}
\vspace{-8mm}
\caption{Ablation Study: latency and area for each optimization applied cumulatively over the baseline, for fastest and smallest Locus PADDs. Annotations show speedups and area reductions with respect to the baseline.}
\label{fig:ablation}
\end{figure}

\begin{figure}[t]
\centering\includegraphics[width=\columnwidth]{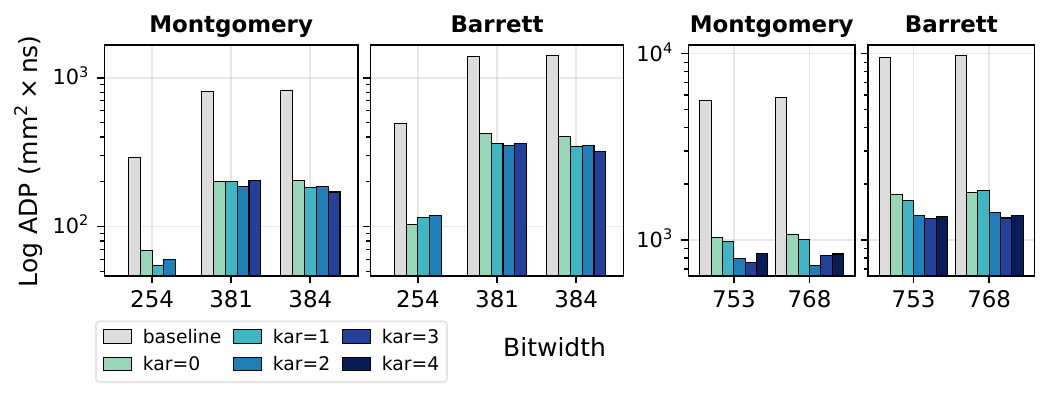}
\vspace{-20pt}
\caption{Area $\times$ Delay across bitwidths for Montgomery \& Barrett PADDs over Karatsuba depths (kar), with all constants variable, with baseline for comparison.}
\label{fig:kar_plots}
\vspace{-5pt}
\end{figure}

\begin{figure}[t]
\centering
\includegraphics[width=\columnwidth]{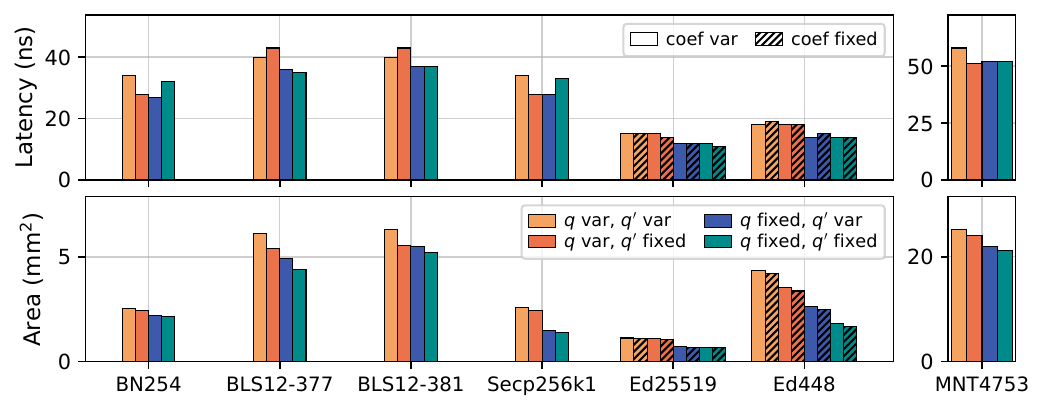}
\vspace{-20pt}
\caption{Fixing Constants in isolation with Montgomery Modmuls and Schoolbook Decomposition.}
\label{fig:fixing_consts_mont}
\vspace{-5pt}
\end{figure}

\subsection{Design Choices and Their Impact}
In this section, we analyze the impact of Locus's key design choices on PADD performance. We begin with an ablation study showing how each optimization cumulatively contributes to the performance of optimal PADDs. We then examine Karatsuba depth and constant-fixing independently, isolating their individual impact. Finally, we compare single- and multi-precision designs.

\subsubsection{Ablation Study} \autoref{fig:ablation} shows the cumulative impact of each optimization on the fastest and smallest Locus PADDs. Optimized multipliers contribute the majority of gains in both latency and area across all curves, achieving roughly $2\times$ speedup and area reduction alone. Constant-fixing provides an additional $\sim$20\% and $\sim$30\% latency and area reduction respectively, but varies by curve.

Certain curves have special-form prime moduli with low Hamming weight. Within our evaluated curves, Secp256k1 and Ed25519 both use pseudo-Mersenne primes \cite{pseudo_mersenne}, and Ed448 \cite{ed448_paper} uses a Solinas prime. Constant-fixing exhibits more pronounced gains for these curves with an additional 40--60\% area reduction for the smallest PADDs, compared to 6--13\% for other curves. Notably, Ed448 achieves up to $3.4\times$ speedup and nearly $8\times$ area reduction over the baseline, the highest among all evaluated curves, highlighting the importance of prime modulus selection in PADD hardware design.

\subsubsection{Impact of Karatsuba Depth} We isolate the effect of Karatsuba depth on PADD performance. \autoref{fig:kar_plots} shows Area Delay Product (ADP) across bitwidths for varying depths, with all constants set to variable and Short Weierstrass formula. Compared to the baseline, designs with our multipliers produce $4\text{-}8\times$ better ADP across all evaluated bitwidths. Lowest ADP typically occurs at intermediate depths (1-3), where it can be up to 30\% better than at the shallowest depth. Non-power-of-2 bitwidths exhibit different optimal depths: Montgomery-based 768-bit designs achieve lowest ADP at $\text{kar}=2$, while 753-bit achieves it at $\text{kar}=3$. This sensitivity is noteworthy as most common named curves operate on non-power-of-2 bitwidths. Barrett continues the trend of performing worse than Montgomery due to its higher computational cost.

\subsubsection{Impact of Fixing Constants} \autoref{fig:fixing_consts_mont} shows latency and area across different constant-fixing combinations and curves, using Montgomery Modmuls and Schoolbook decomposition as our controlled baseline. Our experiments reveal that the relationship between constant-fixing and performance is highly curve-dependent and non-obvious. While fixing all constants consistently produces the smallest area footprint, optimal latency requires nuanced, curve-specific configurations. For instance, BN254 achieves lowest latency when fixing $q$ while keeping $q'$ variable, whereas MNT4753 exhibits the opposite pattern---preferring variable $q$ with fixed $q'$. BLS12-377 shows particularly unusual behavior: despite fully-variable outperforming the variable-$q$, fixed-$q'$ configuration, fixing all constants ultimately yields the lowest latency. Fixing curve coefficients has minimal impact in most cases, though it slightly degrades latency for Ed448. Combined with other design knobs like Karatsuba depth, these interactions become even more complex, making Locus key to uncovering which configurations are optimal.

\subsubsection{Single-precision vs Multi-precision} 
\begin{figure}[t]
\centering\includegraphics[width=\columnwidth]{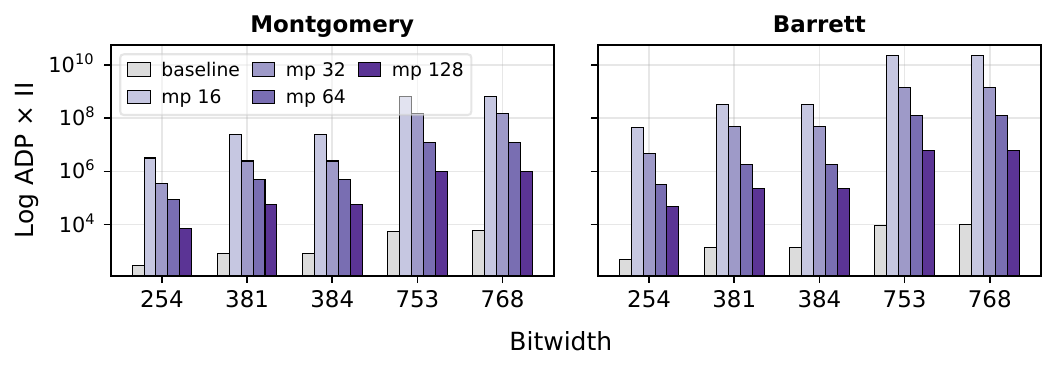}
\vspace{-20pt}
\caption{Area $\times$ Delay $\times$ II product for Montgomery \& Barrett based Multi-Precision PADD designs across word widths with baseline for reference.}
\label{fig:mp_plot}
\vspace{-10pt}
\end{figure}

We use ADP $\times$ II as our comparison metric to fairly account for lower throughput (II $>$ 1) in multi-precision implementations. \autoref{fig:mp_plot} shows multi-precision designs perform significantly worse than even the single-precision baseline across all evaluated bitwidths, but exhibit better ADP $\times$ II as word width increases. While multi-precision achieves much lower area by resource sharing, its latency penalty is severe and is compounded by the inability to fully pipeline.This makes it less suitable for the high-throughput PADD requirements of ZKP accelerators. We note that multi-precision may still be relevant in heavily resource-constrained settings, but for the performance targets of ZKP acceleration, optimized single-precision dominates, motivating our focus on optimizing within the single-precision design space.

\begin{table}[b]
\centering
\vspace{-3mm}
\caption{\small Locus vs. Prior ASICs}
\vspace{-3mm}
\label{tab:padd_asic_prior}
\footnotesize
\resizebox{1\columnwidth}{!}{
\setlength{\tabcolsep}{2mm}{
\begin{tabular}{|c|c|c|c|c|c|}
\hline
\textbf{Curve} & \textbf{Design} & \textbf{Freq.} & \textbf{\begin{tabular}[c]{@{}c@{}}Latency\\ (ns)\end{tabular}} & \textbf{\begin{tabular}[c]{@{}c@{}}Area\\ (mm$^2$)\end{tabular}} & \textbf{\begin{tabular}[c]{@{}c@{}}Power\\ (W)\end{tabular}} \\ \hline
\multirow{3}{*}{\textbf{BN254}} 
& Baseline HLS & 1 GHz & 63 & 5.10 & 4.72 \\ \cline{2-6}
& Locus Fastest & 1 GHz & 27 (\textbf{2.33$\times$}) & 2.21 \textbf{(2.31$\times$)} & 5.13 \\ \cline{2-6}
& Locus Smallest & 1 GHz & 35 (\textbf{1.80$\times$}) & 1.76 \textbf{(2.90$\times$)} & 3.64 \\ \hline 
\hline
\multirow{3}{*}{\textbf{BLS12-381}} 
& Baseline HLS & 1 GHz & 82 & 11.94 & 8.26 \\ \cline{2-6}
& Locus Fastest & 1 GHz & 31 (\textbf{2.65$\times$}) & 4.65 \textbf{(2.57$\times$)} & 6.83 \\ \cline{2-6}
& Locus Smallest & 1 GHz & 49 (\textbf{1.67$\times$}) & 3.88 \textbf{(3.08$\times$)} & 7.11 \\ \hline 
\hline
\multirow{3}{*}{\textbf{BLS12-377}} 
& Baseline HLS & 1 GHz & 82 & 11.70 & 8.12 \\ \cline{2-6}
& Locus Fastest & 1 GHz & 31 (\textbf{2.65$\times$}) & 3.88 \textbf{(3.01$\times$)} & 5.87 \\ \cline{2-6}
& Locus Smallest & 1 GHz & 44 (\textbf{1.86$\times$}) & 3.54 \textbf{(3.30$\times$)} & 7.15 \\ \hline 
\hline
\multirow{3}{*}{\textbf{MNT4753}} 
& Baseline HLS & 667 MHz & 169.5 & 45.89 & 21.76 \\ \cline{2-6}
& Locus Fastest & 1 GHz & 51 (\textbf{3.32$\times$}) & 24.00 \textbf{(1.91$\times$)} & 44.16 \\ \cline{2-6}
& Locus Smallest & 1 GHz & 68 (\textbf{2.49$\times$}) & 14.43 \textbf{(3.18$\times$)} & 23.04 \\ \hline
\end{tabular}
}
}
\end{table}

\begin{table}[t]
\centering
\caption{\small Locus vs. Prior FPGA - BLS12-377*}
\vspace{-8pt}
\label{tab:fpga_comparison}
\footnotesize
\resizebox{1\columnwidth}{!}{
\setlength{\tabcolsep}{3pt}{
\begin{tabular}{|>{\centering\arraybackslash}p{0.2475\columnwidth}|c|c|c|c|c|>{\centering\arraybackslash}p{0.08\columnwidth}|>{\centering\arraybackslash}p{0.1\columnwidth}|}
\hline
\textbf{Design} & \textbf{Model} & \textbf{LUTs} & \textbf{Regs} & \textbf{DSPs} & \textbf{Cycles} & \textbf{\begin{tabular}[c]{@{}c@{}}Fmax\\ (MHz)\end{tabular}} & \textbf{\begin{tabular}[c]{@{}c@{}}Latency\\ (\textmu s)\end{tabular}} \\
\hline
\textbf{Locus Fastest} & VU9P & 313,438 & 262,880 & 4,361 & 50 & 291 & 0.172 \\
\hline
\textbf{Locus Min-DSP} & VU9P & 498,716 & 369,423 & 2,436 & 52 & 259 & 0.201 \\
\hline
\textbf{Locus Balanced} & VU9P & 416,597 & 289,739 & 3,402 & 53 & 277 & 0.191 \\
\hline
\textbf{Locus High-Fmax} & VU9P & 505,704 & 437,901 & 4,032 & 75 & 408 & 0.184 \\
\hline
\hline 
\textbf{Locus Fastest} & VH1782 & 344,658 & 296,964 & 4,592 & 51 & 308 & 0.166 \\
\hline
\textbf{Locus Min-DSP} & VH1782 & 569,452 & 417,245 & 1,512 & 69 & 241 & 0.286 \\
\hline
\textbf{Locus Balanced} & VH1782 & 355,861 & 289,557 & 2,898 & 65 & 279 & 0.233 \\
\hline \hline 
\textbf{CycloneMSM} \cite{cyclonemsm} & VU9P & 310,717 & 337,944 & 2,268 & 96 & 250 & 0.384 \\
\hline
\textbf{HardcamlMSM} \cite{hardcamlmsm} & VU9P & -- & -- & -- & 238 & 278 & 0.856 \\
\hline
\textbf{BSTMSM} \cite{bstmsm} & U250 & -- & -- & -- & 260 & 300 & 0.867 \\
\hline
\end{tabular}
}
}
\vspace{-3mm}
\end{table}

\subsection{Locus PADDs vs. Prior Work}
\subsubsection{ASIC} Using Locus, we generate PADDs following the approach of prior works \cite{szkp, zkspeed} (baseline multipliers with constant-fixing disabled) to compare with our corresponding best performant designs (\autoref{tab:padd_asic_prior}). We find that our fastest designs achieve a $2.71\times$ geomean speedup, while our smallest designs achieve a $3.11\times$ geomean area reduction over prior works. Our power and power density estimates are consistent with prior works \cite{deepraj1}.

\subsubsection{FPGA}
Prior works on FPGA \cite{cyclonemsm, hardcamlmsm, bstmsm} largely implement BLS12-377*. For comparison, we implement CycloneMSM's PADD formula. We select designs targeting different objectives: matching prior work clock speeds, minimizing DSPs (a critical resource constraint), and balanced designs (\autoref{tab:fpga_comparison}).

On the VU9P FPGA, Locus designs achieve $2$-$5\times$ speedup with $1.3$-$1.9\times$ fewer cycles than CycloneMSM and $3$-$5\times$ fewer cycles than HardcamlMSM and BSTMSM. Locus produces designs that achieve clock speeds as high as 408 MHz in just 75 cycles. Our min-DSP design uses only 7.4\% more DSPs than CycloneMSM, while delivering nearly $2\times$ speedup, though with higher LUT and register usage.
On the state-of-the-art VH1782 FPGA, our min-DSP variant uses only 1,512 DSPs---the lowest among implementations with reported DSP counts---while maintaining low cycles and latencies.

While Locus designs use more LUTs, prior works generally target a single implementation, whereas Locus generates a design space spanning diverse resource--performance trade-offs, enabling designers to select configurations that best match their constraints.

\begin{table}[t]
\centering
\caption{\small Latency (ns) and Speedup of Locus vs CPU}
\vspace{-1em}
\label{tab:padd_locus_vs_cpu}
\small
\resizebox{1\columnwidth}{!}{
\setlength{\tabcolsep}{2mm}{
\begin{tabular}{|c|c|c|c|}
\hline
\textbf{Curve} & \textbf{\begin{tabular}[c]{@{}c@{}}CPU\\ {1 Thread}\end{tabular}} & \textbf{\begin{tabular}[c]{@{}c@{}}Locus Fastest FPGA \\(Speedup vs CPU)\end{tabular}} & \textbf{\begin{tabular}[c]{@{}c@{}}Locus Fastest ASIC\\(Speedup vs CPU/FPGA)\end{tabular}} \\ \hline
\textbf{BN254} & 522.28 & 431.35 \textbf{(1.21$\times$)} & 27.00 \textbf{(19.34$\times$ / 15.98$\times$)}  \\ \hline
\textbf{BLS12-381} & 1019.86 & 558.49 \textbf{(1.83$\times$)} & 31.00 \textbf{(32.90$\times$ / 18.02$\times$)} \\ \hline
\textbf{BLS12-377} & 998.57 & 561.74 \textbf{(1.78$\times$)} & 31.00 \textbf{(32.21$\times$ / 18.12$\times$)} \\ \hline
\textbf{MNT4753} & 3597.22 & -- & 51.00 \textbf{(70.53$\times$ / --)} \\ \hline
\end{tabular}
}
}
\vspace{-3mm}
\end{table}
\begin{table}[t]
\centering
\footnotesize
\caption{\small MSM Design Space}
\label{tab:design_space_msm}
\vspace{-4mm}
{
\begin{tabular}{|c|c|}
\hline
\textbf{Design Setting} & \textbf{Values} \\ \hline \hline
Scalar Width ($\lambda$) & 254 (BN254), 253 (BLS12-377) \\
& 255 (BLS12-381), 753 (MNT4753) \\ \hline
PE Count ($K$) & 1, 2, 4, 8, 16 \\ \hline
Window size ($w$) & $5-20$ \\ \hline
Words on-chip ($P'$) & 1K, 2K, $\ldots$ 256K \\ \hline
\end{tabular}
}
\vspace{-3mm}
\end{table}


\subsubsection{Speedup over CPU}
In \autoref{tab:padd_locus_vs_cpu}, we compare the latencies of our fastest VH1782 FPGA and ASIC designs against a single-threaded AMD EPYC 7502 CPU, using point-addition implementations from the arkworks \cite{arkworks} library, a widely used and optimized Rust library for finite-field and elliptic-curve arithmetic used in ZKP protocols.
We evaluate major ZKP curves (MNT4753 designs are too large to fit on our FPGAs); our FPGA designs achieve 1.21-1.83$\times$ speedups, and our ASIC designs achieve 19-70$\times$ and 15-18$\times$ speedup over CPU and their FPGA counterparts, respectively. 

\begin{figure}[t]
\centering
\includegraphics[width=0.95\columnwidth]{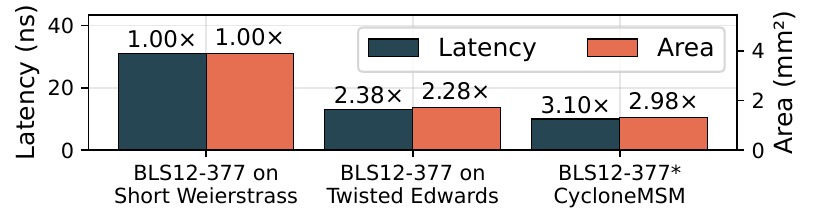}
\vspace{-10pt}
\caption{Comparison of fastest BLS12-377 designs for different PADD formulas on ASIC. Annotations show speedup and area reduction compared to BLS12-377 on Short Weierstrass.}
\label{fig:bls12_377_comparison}
\vspace{-2mm}
\end{figure}

\begin{figure}[t]
\centering
\vspace{-2mm}
\includegraphics[width=\columnwidth]{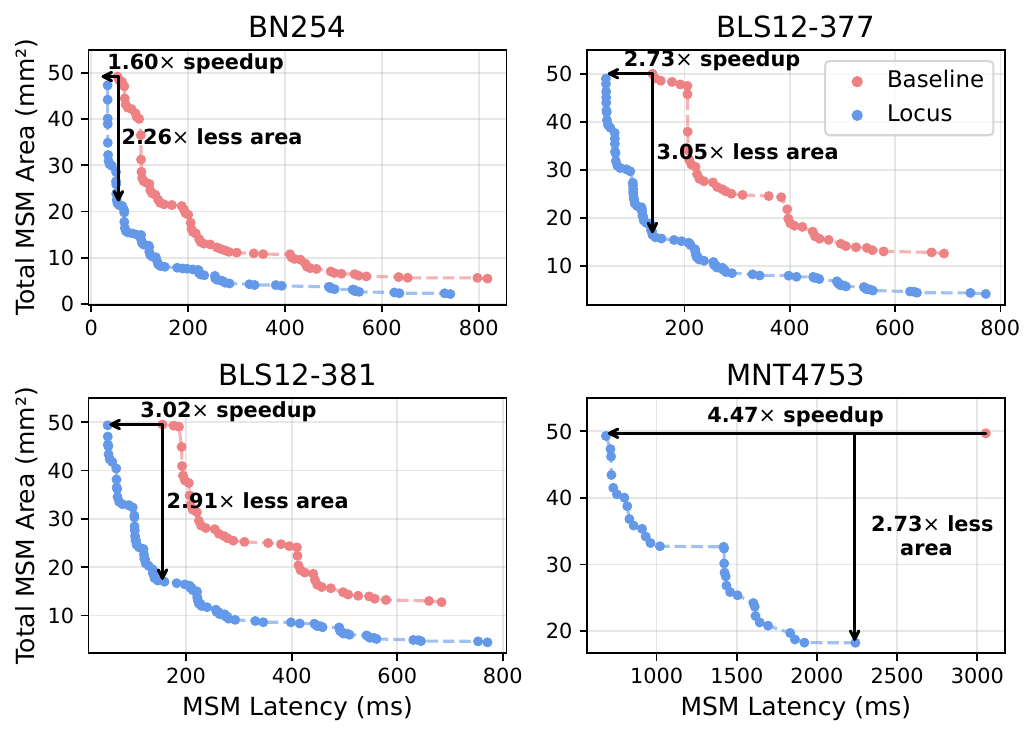}
\vspace{-6mm}
\caption{Pareto-optimality space of our MSMs versus baseline designs in a 50 mm$^2$ area budget. Horizontal arrows are annotated with the iso-area speedup compared to fastest baselines. Vertical arrows are annotated with the iso-latency (where possible) area reduction compared to fastest baselines.}
\label{fig:msm_pareto_UPDATED}
\end{figure}

\subsubsection{Rapid Algorithm Exploration} CycloneMSM achieves exceptional PADD performance by optimizing the PADD formula for BLS12-377*, specifically eliminating a key Modmul bottleneck.
Using Locus, we generate this design on ASIC, and find that
CycloneMSM's PADD achieves $3.10\times$ speedup and $2.98\times$ area reduction versus standard Short Weierstrass (\autoref{fig:bls12_377_comparison}),
achieving a very low 10 ns latency. This demonstrates Locus's ability to rapidly explore algorithmic optimizations and assess hardware metrics.

\begin{table}[t]
\centering
\caption{\small Optimal MSM configurations using Baseline and Locus PADD designs. Total workload length is $2^{24}$, and $P'$ refers to the number of points stored on-chip.}
\vspace{-2mm}
\label{tab:msm_config}
\resizebox{1\columnwidth}{!}{
\begin{tabular}{|c|ccccc|ccccc|}
\hline
\multirow{2}{*}{\textbf{Curve}} 
 & \multicolumn{5}{c|}{\textbf{Fastest Baseline Design}} & \multicolumn{5}{c|}{\textbf{Fastest Locus Design}} \\  \cline{2-11}
 & \multicolumn{1}{c|}{$w^{*}$} & \multicolumn{1}{c|}{$K$} & \multicolumn{1}{c|}{$P'$} & \multicolumn{1}{c|}{$t_{add}$} & $t_{dbl}$ & \multicolumn{1}{c|}{$w^{*}$} & \multicolumn{1}{c|}{$K$} & \multicolumn{1}{c|}{$P'$} & \multicolumn{1}{c|}{$t_{add}$} & $t_{dbl}$ \\ \hline
\textbf{BN254} & \multicolumn{1}{c|}{11} & \multicolumn{1}{c|}{8} & \multicolumn{1}{c|}{8K} & \multicolumn{1}{c|}{63} & 32 & \multicolumn{1}{c|}{8} & \multicolumn{1}{c|}{16} & \multicolumn{1}{c|}{64K} & \multicolumn{1}{c|}{27} & 14 \\ \hline
\textbf{BLS12-377} & \multicolumn{1}{c|}{8} & \multicolumn{1}{c|}{4} & \multicolumn{1}{c|}{8K} & \multicolumn{1}{c|}{82} & 41 & \multicolumn{1}{c|}{11} & \multicolumn{1}{c|}{8} & \multicolumn{1}{c|}{32K} & \multicolumn{1}{c|}{31} & 16 \\ \hline
\textbf{BLS12-381} & \multicolumn{1}{c|}{8} & \multicolumn{1}{c|}{4} & \multicolumn{1}{c|}{2K} & \multicolumn{1}{c|}{82} & 41 & \multicolumn{1}{c|}{11} & \multicolumn{1}{c|}{8} & \multicolumn{1}{c|}{32K} & \multicolumn{1}{c|}{31} & 16 \\ \hline
\textbf{MNT4753} & \multicolumn{1}{c|}{7} & \multicolumn{1}{c|}{1} & \multicolumn{1}{c|}{1K} & \multicolumn{1}{c|}{113} & 56 & \multicolumn{1}{c|}{10} & \multicolumn{1}{c|}{2} & \multicolumn{1}{c|}{2K} & \multicolumn{1}{c|}{58} & 29 \\ \hline
\end{tabular}
}
\vspace{-5mm}
\end{table}

\begin{table*}[t]
\centering

\begin{minipage}[t]{0.32\textwidth}
\centering
\captionof{table}{\small PADD Configurations for \ End-to-End Analysis}
\label{tab:e2e_knobs}
\vspace{-3.5mm}

\resizebox{\linewidth}{!}{
\begin{tabular}{|c|c|c|c|c|}
\hline
\textbf{Design} &
\textbf{\begin{tabular}[c]{@{}c@{}}Tech\\Node\end{tabular}} &
\textbf{Freq.} &
\textbf{\begin{tabular}[c]{@{}c@{}}PADD\\Stages\end{tabular}} &
\textbf{Approach} \\ \hline
SZKP     & 22nm & 300 MHz & 38  & Baseline HLS \\ \hline \hline
A & 12nm & 667 MHz & 113 & Baseline HLS \\ \hline
B & 12nm & 667 MHz & 34  & Locus \\ \hline
C & 12nm & 1 GHz   & 51  & Locus \\ \hline \hline
D & 12nm & 667 MHz & 113 & Baseline HLS \\ \hline
E & 12nm & 667 MHz & 34  & Locus \\ \hline
F & 12nm & 1 GHz   & 51  & Locus \\ \hline
G & 12nm & 1 GHz   & 51  & Locus (2 PADD) \\ \hline
\end{tabular}
}
\end{minipage}
\hfill
\begin{minipage}[t]{0.67\textwidth}
\centering
\captionof{table}{\small Groth16 Runtime (ms) on MNT4753. Speedups are relative to \newline Design A for 5-bit windows, Design D for 7-bit windows.}
\label{tab:e2e_zkp}
\vspace{-2mm}

\resizebox{\linewidth}{!}{
\begin{tabular}{|c|c|c|c|c|c||c|c|c|c|}
\hline
\multirow{2}{*}{\textbf{Workload}} & \multirow{2}{*}{\textbf{Size}} &
\multicolumn{4}{c||}{\textbf{Designs with Window Size = 5}} &
\multicolumn{4}{c|}{\textbf{Designs with Window Size = 7}} \\ \cline{3-10}
& & \textbf{SZKP} & \textbf{A} & \textbf{B} & \textbf{C} &
\textbf{D} & \textbf{E} & \textbf{F} & \textbf{G} \\ \hline \hline

\textbf{AES} & 16384 & 17.20 & 15.89 & 4.99 & 4.89 \textbf{(3.25$\times$)} & 8.19 & 4.50 & 3.52 \textbf{(2.33$\times$)} & 1.96 \textbf{(4.19$\times$)} \\ \hline
\textbf{SHA2} & 32768 & 33.09 & 30.05 & 9.49 & 9.28 \textbf{(3.24$\times$)} & 11.65 & 7.60 & 5.64 \textbf{(2.07$\times$)} & 3.19 \textbf{(3.65$\times$)} \\ \hline
\textbf{RSA} & 98304 & 98.09 & 87.53 & 28.37 & 27.42 \textbf{(3.19$\times$)} & 26.34 & 20.90 & 14.67 \textbf{(1.79$\times$)} & 8.72 \textbf{(3.02$\times$)} \\ \hline
\textbf{RSASigVer} & 131072 & 128.30 & 114.88 & 36.59 & 35.65 \textbf{(3.22$\times$)} & 32.30 & 26.33 & 18.36 \textbf{(1.76$\times$)} & 10.65 \textbf{(3.03$\times$)} \\ \hline
\textbf{MerkleTree} & 294912 & 300.88 & 262.87 & 88.20 & 83.91 \textbf{(3.13$\times$)} & 73.32 & 63.98 & 43.87 \textbf{(1.67$\times$)} & 27.39 \textbf{(2.68$\times$)} \\ \hline
\textbf{Auction} & 557056 & 572.20 & 496.99 & 168.70 & 159.65 \textbf{(3.11$\times$)} & 136.27 & 122.16 & 83.23 \textbf{(1.64$\times$)} & 52.70 \textbf{(2.59$\times$)} \\ \hline \hline

\multicolumn{2}{|c|}{\textbf{Chip Area (mm$^2$)}} &
91.00 & 65.08 & 23.10 & 34.97 & 67.27 & 25.29 & 37.16 & 61.23 \\ \hline
\end{tabular}
}

\end{minipage}
\vspace{-3mm}
\end{table*}

\subsection{Locus MSMs vs. Prior Work}
We now evaluate Locus PADDs in end-to-end MSM implementations. 
We study four curves used in prior ZKP ASICs \cite{szkp, zkspeed, pipezk, zkphire, priormsm} using the latency model from Algorithm \ref{alg:msm} and the areas of synthesized PADDs and memory in our cost model.
We first extract the Pareto-optimal PADD designs for each curve from \autoref{fig:design_space_padd}. Then, for each Pareto-optimal PADD, we sweep all MSM configurations from \autoref{tab:design_space_msm} and construct the MSM-level Pareto curves (\autoref{fig:msm_pareto_UPDATED}) assuming a $50$ mm$^2$ area budget. 
We validate the predicted latencies with cycle-accurate simulation of the SZKP architecture.

For fair comparison with prior works (zkSpeed and SZKP), we implement their PADDs (i.e., Catapult-generated without Locus optimizations), targeting the same technology node we use. 
From \autoref{fig:msm_pareto_UPDATED} we see that across all curves, the fastest baseline MSMs are $2-3\times$ larger than equivalent Locus designs at iso-latency.
For context, zkSpeed's MSM compute area is $105\,\text{mm}^2$. A $3\times$ reduction would shrink this to $35\,\text{mm}^2$, lowering their total compute area from $163$ to $94$ mm$^2$, a $42\%$ overall reduction.
Compared to iso-area, Locus designs achieve geomean $2.72\times$ speedup. The speedup gains are more pronounced at higher bitwidths (e.g. for MNT4753) because baseline designs cannot clock at 1 GHz; Locus's approach enables these bitwidths to clock at 1 GHz.
Comparing the fastest Locus designs with the fastest baseline designs, Locus achieves roughly $2-4\times$ improvement in performance per area across curves.
This is because Locus’s optimizations yield faster, smaller PADDs as seen in \autoref{tab:msm_config}. This allows for more MSM PEs and additional on-chip memory to reduce pipeline fill/drain penalties.
As a result, Locus improves both PADD efficiency and end-to-end MSM performance.

\subsection{End-to-End ZKP Evaluation}

To evaluate the system-level impact of Locus PADDs beyond MSM, we integrate our optimal PADD designs into an SZKP-style architecture focusing on the NTT and dense MSM critical path for Groth16. SZKP's original evaluation targets TSMC 22nm technology node with the MNT4753 curve, using MSM window size $w=5$ and a 38-stage PADD pipeline clocked at 300 MHz. Since our ASIC evaluation targets GF 12nm, we progressively modify the design one parameter at a time, as summarized in \autoref{tab:e2e_knobs}, to separate the effects of technology scaling, Locus optimization, frequency scaling, window size, and additional PADD parallelism.
We also use Locus-optimized Modmuls in NTT butterflies, using the same design as SZKP. We then use cycle-accurate simulation to determine end-to-end runtime; runtimes and area are reported in \autoref{tab:e2e_zkp}.

We first consider $w=5$. Design A ports SZKP's straight-line HLS design to 12nm, where the PADD reaches a maximum frequency of 667 MHz with a 113-stage pipeline, nearly $3\times$ deeper than the original 38-stage design at 300 MHz. At $w=5$, there are only $2^5-1=31$ buckets, far fewer than the 113 pipeline stages, preventing full PADD utilization.
With Locus optimizations at iso-frequency (Design B), we obtain a 34-stage PADD, improving utilization and reducing the runtime considerably. Tuning up the frequency to 1 GHz (Design C), Locus-optimized PADDs still meet timing, though the PADD depth extends to 51 stages. While this represents a lower-utilization MSM architecture than with a 34-stage PADD, the frequency increase is enough to yield an overall reduced runtime, at a higher area cost. As such, Design C achieves $3.19\times$ geomean speedup over Design A.

To achieve better utilization (and a lower PADD op count), we can increase the window size to $w=7$ with $2^7 - 1 = 127$ buckets. This does not affect the PADD area, but does increase the memory requirement slightly (bucket accumulation registers and number of queues). Consequently, we can see Design D is slightly larger than Design A, but the PADD depth is less than the number of buckets, significantly improving utilization and reducing runtime. We observe similar trends for Designs E-G, except that transitioning from $E \rightarrow F$ exhibits greater improvement because the PADD depth is still much less than bucket count. With Locus's area-saving optimizations, we can include a second PADD in Design G, achieving geomean speedup of $3.15\times$ over Design D with less area. 

These results highlight a broader architectural insight: SZKP's simple scheduling schemes (e.g., round-robin, longest-queue, etc.) rely on the pipeline depth being short relative to the number of buckets in order to achieve near-perfect utilization. As we scale to higher frequencies and larger bitwidths, PADDs require increasingly deep pipelines that violate this assumption, degrading utilization and limiting the benefits of technology scaling. Locus PADDs tame this pipeline depth at 1 GHz, preserving SZKP's simple scheduling without requiring more complex schedulers like those in \cite{pipezk, priormsm, gypsophila}. Therefore, Locus not only improves PADD-level metrics but also enables the overall accelerator architecture to scale.
\vspace{-3mm}
\section{Conclusion}

We present Locus, a framework for exploring and optimizing elliptic-curve point addition hardware. Using Locus, we perform the first comprehensive hardware-focused exploration of the PADD design space across curves used in ZKPs, blockchains, and digital signatures, spanning 1,000+ designs and exposing key architectural trade-offs. Locus rapid generates and evaluates PADD implementations tailored to performance and area targets across ASICs and FPGAs. We integrate optimized PADDs into an MSM and a full ZKP accelerator, demonstrating gains at both the MSM and proof-generation levels. Looking forward, Locus can be extended to other cryptographic modules using its optimized modular-arithmetic units.

\vspace{-3mm}
\begin{acks}
This work was supported by NSF CAREER award
\#2340137, NSF CIRC GRAND \#2450539, and NSF NeTs
\#2504400, and generous support from DTCC, AMD and
Google. The views, opinions, and/or findings expressed are
those of the authors and do not necessarily reflect the views of sponsors.
\end{acks}

\balance
\bibliographystyle{ACM-Reference-Format}
\bibliography{refs}

@String{Computing = "Computing" }

@String{Computer = "{IEEE} Computer" }

@String{Springer = "Springer-Verlag" }

@misc{cyclonemsm,
  author       = {Kaveh Aasaraai and Don Beaver and Emanuele Cesena and Rahul Maganti and Nicolas Stalder and Javier Varela},
  title        = {{FPGA} Acceleration of Multi-Scalar Multiplication: {CycloneMSM}},
  howpublished = {Cryptology ePrint Archive, Paper 2022/1396},
  year         = {2022},
  url          = {https://eprint.iacr.org/2022/1396}
}

@inproceedings{hardcamlmsm,
author = {Ray, Andy and Devlin, Benjamin and Quah, Fu Yong and Yesantharao, Rahul},
title = {Hardcaml MSM: A High-Performance Split CPU-FPGA Multi-Scalar Multiplication Engine},
year = {2024},
isbn = {9798400704185},
publisher = {Association for Computing Machinery},
address = {New York, NY, USA},
url = {https://doi.org/10.1145/3626202.3637577},
doi = {10.1145/3626202.3637577},
booktitle = {Proceedings of the 2024 ACM/SIGDA International Symposium on Field Programmable Gate Arrays},
pages = {33–39},
numpages = {7},
location = {Monterey, CA, USA},
series = {FPGA '24}
}

@article{karatsuba1995complexity,
  title={The complexity of computations},
  author={Karatsuba, Anatolii Alexeevich},
  journal={Proceedings of the Steklov Institute of Mathematics-Interperiodica Translation},
  volume={211},
  pages={169--183},
  year={1995},
  publisher={Providence, RI: American Mathematical Society}
}

@INPROCEEDINGS{bstmsm,
  author={Zhao, Baoze and Huang, Wenjin and Li, Tianrui and Huang, Yihua},
  booktitle={2023 International Conference on Field Programmable Technology (ICFPT)}, 
  title={BSTMSM: A High-Performance FPGA-based Multi-Scalar Multiplication Hardware Accelerator}, 
  year={2023},
  volume={},
  number={},
  pages={35-43},
  doi={10.1109/ICFPT59805.2023.00009}}

@inproceedings{msmac,
author = {Qiu, Pengcheng and Wu, Guiming and Chu, Tingqiang and Wei, Changzheng and Luo, Runzhou and Yan, Ying and Wang, Wei and Zhang, Hui},
title = {MSMAC: Accelerating Multi-Scalar Multiplication for Zero-Knowledge Proof},
year = {2024},
isbn = {9798400706011},
publisher = {Association for Computing Machinery},
address = {New York, NY, USA},
url = {https://doi.org/10.1145/3649329.3655672},
doi = {10.1145/3649329.3655672},
booktitle = {Proceedings of the 61st ACM/IEEE Design Automation Conference},
articleno = {66},
numpages = {6},
location = {San Francisco, CA, USA},
series = {DAC '24}
}

@inproceedings{szkp, series={PACT ’24},
   title={SZKP: A Scalable Accelerator Architecture for Zero-Knowledge Proofs},
   url={http://dx.doi.org/10.1145/3656019.3676898},
   DOI={10.1145/3656019.3676898},
   booktitle={Proceedings of the 2024 International Conference on Parallel Architectures and Compilation Techniques},
   publisher={ACM},
   author={Daftardar, Alhad and Reagen, Brandon and Garg, Siddharth},
   year={2024},
   month=oct, pages={271–283},
   collection={PACT ’24} }

@inproceedings{mo2025mtu,
  title={Mtu: The multifunction tree unit for accelerating zero-knowledge proofs},
  author={Mo, Jianqiao and Daftardar, Alhad and Ah-Kiow, Joey and Guo, Kaiyue and B{\"u}nz, Benedikt and Garg, Siddharth and Reagen, Brandon},
  booktitle={Proceedings of the 14th International Workshop on Hardware and Architectural Support for Security and Privacy},
  pages={19--27},
  year={2025}
}

@article{priormsm,
author = {Liu, Changxu and Zhou, Hao and Dai, Patrick and Shang, Li and Yang, Fan},
title = {PriorMSM: An Efficient Acceleration Architecture for Multi-Scalar Multiplication},
year = {2024},
issue_date = {September 2024},
publisher = {Association for Computing Machinery},
address = {New York, NY, USA},
volume = {29},
number = {5},
issn = {1084-4309},
url = {https://doi.org/10.1145/3678006},
doi = {10.1145/3678006},
journal = {ACM Trans. Des. Autom. Electron. Syst.},
month = aug,
articleno = {77},
numpages = {26}
}

@article{hyperelliptic,
  title={Explicit-formulas database},
  author={Bernstein, Daniel J},
  url={http://www.hyperelliptic.org/EFD},
  year={2007}
}

@inproceedings{pipezk,
author = {Zhang, Ye and Wang, Shuo and Zhang, Xian and Dong, Jiangbin and Mao, Xingzhong and Long, Fan and Wang, Cong and Zhou, Dong and Gao, Mingyu and Sun, Guangyu},
title = {PipeZK: accelerating zero-knowledge proof with a pipelined architecture},
year = {2021},
isbn = {9781450390866},
publisher = {IEEE Press},
url = {https://doi.org/10.1109/ISCA52012.2021.00040},
doi = {10.1109/ISCA52012.2021.00040},
booktitle = {Proceedings of the 48th Annual International Symposium on Computer Architecture},
pages = {416–428},
numpages = {13},
location = {Virtual Event, Spain},
series = {ISCA '21}
}

@inproceedings{zkspeed,
author = {Daftardar, Alhad and Mo, Jianqiao and Ah-kiow, Joey and B\"{u}nz, Benedikt and Karri, Ramesh and Garg, Siddharth and Reagen, Brandon},
title = {Need for zkSpeed: Accelerating HyperPlonk for Zero-Knowledge Proofs},
year = {2025},
isbn = {9798400712616},
publisher = {Association for Computing Machinery},
address = {New York, NY, USA},
url = {https://doi.org/10.1145/3695053.3731021},
doi = {10.1145/3695053.3731021},
booktitle = {Proceedings of the 52nd Annual International Symposium on Computer Architecture},
pages = {1986–2001},
numpages = {16},
location = {
},
series = {ISCA '25}
}

@INPROCEEDINGS{zkphire,
  author={Daftardar, Alhad and Mo, Jianqiao and Ah-kiow, Joey and Bünz, Benedikt and Garg, Siddharth and Reagen, Brandon},
  booktitle={2026 IEEE International Symposium on High Performance Computer Architecture (HPCA)}, 
  title={zkPHIRE: A Programmable Accelerator for ZKPs over HIgh-degRee, Expressive Gates}, 
  year={2026},
  volume={},
  number={},
  pages={1-15},
  doi={10.1109/HPCA68181.2026.11408480}}

@article{montgomery,
 ISSN = {00255718, 10886842},
 URL = {http://www.jstor.org/stable/2007970},
 author = {Peter L. Montgomery},
 journal = {Mathematics of Computation},
 number = {170},
 pages = {519--521},
 publisher = {American Mathematical Society},
 title = {Modular Multiplication Without Trial Division},
 urldate = {2025-11-13},
 volume = {44},
 year = {1985}
}

@inproceedings{barrett,
  title={Implementing the Rivest Shamir and Adleman Public Key Encryption Algorithm on a Standard Digital Signal Processor},
  booktitle={Advances in Cryptology - CRYPTO '86, Santa Barbara, California, USA, 1986, Proceedings},
  series={Lecture Notes in Computer Science},
  publisher={Springer},
  volume={263},
  pages={311-323},
  doi={10.1007/3-540-47721-7_24},
  author={Paul Barrett},
  year=1986
}

@inproceedings{zkp,
author = {Blum, Manuel and Feldman, Paul and Micali, Silvio},
title = {Non-interactive zero-knowledge and its applications},
year = {1988},
isbn = {0897912640},
publisher = {Association for Computing Machinery},
address = {New York, NY, USA},
url = {https://doi.org/10.1145/62212.62222},
doi = {10.1145/62212.62222},
booktitle = {Proceedings of the Twentieth Annual ACM Symposium on Theory of Computing},
pages = {103–112},
numpages = {10},
location = {Chicago, Illinois, USA},
series = {STOC '88}
}

@inproceedings{gypsophila,
author = {Liu, Changxu and Zhou, Hao and Yang, Lan and Xu, Jiamin and Dai, Patrick and Yang, Fan},
title = {Gypsophila: A Scalable and Bandwidth-Optimized Multi-Scalar Multiplication Architecture},
year = {2024},
isbn = {9798400706011},
publisher = {Association for Computing Machinery},
address = {New York, NY, USA},
url = {https://doi.org/10.1145/3649329.3658259},
doi = {10.1145/3649329.3658259},
booktitle = {Proceedings of the 61st ACM/IEEE Design Automation Conference},
articleno = {94},
numpages = {6},
location = {San Francisco, CA, USA},
series = {DAC '24}
}

@book{mp_src,
author = {Menezes, Alfred J. and Vanstone, Scott A. and Oorschot, Paul C. Van},
title = {Handbook of Applied Cryptography},
year = {1996},
isbn = {0849385237},
publisher = {CRC Press, Inc.},
address = {USA},
edition = {1st}
}

@book{gmp,
author = {Granlund, Torbjrn and Gmp Development Team},
title = {GNU MP 6.0 Multiple Precision Arithmetic Library},
year = {2015},
isbn = {9789888381968},
publisher = {Samurai Media Limited},
address = {London, GBR}
}

@inproceedings{zhang_cpu,
author = {Zhang, Naifeng and Fu, Sophia and Franchetti, Franz},
title = {Towards Closing the Performance Gap for Cryptographic Kernels Between CPUs and Specialized Hardware},
year = {2025},
isbn = {9798400715730},
publisher = {Association for Computing Machinery},
address = {New York, NY, USA},
url = {https://doi.org/10.1145/3725843.3756120},
doi = {10.1145/3725843.3756120},
booktitle = {Proceedings of the 58th IEEE/ACM International Symposium on Microarchitecture},
pages = {1704–1718},
numpages = {15},
location = {
},
series = {MICRO '25}
}

@inproceedings{zhang_gpu,
author = {Zhang, Naifeng and Franchetti, Franz},
title = {Code Generation for Cryptographic Kernels using Multi-word Modular Arithmetic on GPU},
year = {2025},
isbn = {9798400712753},
publisher = {Association for Computing Machinery},
address = {New York, NY, USA},
url = {https://doi.org/10.1145/3696443.3708948},
doi = {10.1145/3696443.3708948},
booktitle = {Proceedings of the 23rd ACM/IEEE International Symposium on Code Generation and Optimization},
pages = {476–492},
numpages = {17},
location = {Las Vegas, NV, USA},
series = {CGO '25}
}

@misc{groth,
      author = {Jens Groth},
      title = {On the Size of Pairing-based Non-interactive Arguments},
      howpublished = {Cryptology ePrint Archive, Paper 2016/260},
      year = {2016},
      url = {https://eprint.iacr.org/2016/260}
}

@misc{hyperplonk,
      author = {Binyi Chen and Benedikt Bünz and Dan Boneh and Zhenfei Zhang},
      title = {{HyperPlonk}: Plonk with Linear-Time Prover and High-Degree Custom Gates},
      howpublished = {Cryptology {ePrint} Archive, Paper 2022/1355},
      year = {2022},
      url = {https://eprint.iacr.org/2022/1355}
}

@inproceedings{orion,
  author       = {Tiancheng Xie and
                  Yupeng Zhang and
                  Dawn Song},
  editor       = {Yevgeniy Dodis and
                  Thomas Shrimpton},
  title        = {Orion: Zero Knowledge Proof with Linear Prover Time},
  booktitle    = {Advances in Cryptology - {CRYPTO} 2022 - 42nd Annual International
                  Cryptology Conference, {CRYPTO} 2022, Santa Barbara, CA, USA, August
                  15-18, 2022, Proceedings, Part {IV}},
  series       = {Lecture Notes in Computer Science},
  volume       = {13510},
  pages        = {299--328},
  publisher    = {Springer},
  year         = {2022},
  url          = {https://doi.org/10.1007/978-3-031-15985-5\_11},
  doi          = {10.1007/978-3-031-15985-5\_11},
  bibsource    = {dblp computer science bibliography, https://dblp.org}
}

@INPROCEEDINGS{pippenger,
  author={Pippenger, Nicholas},
  booktitle={17th Annual Symposium on Foundations of Computer Science (sfcs 1976)}, 
  title={On the evaluation of powers and related problems}, 
  year={1976},
  volume={},
  number={},
  pages={258-263},
  doi={10.1109/SFCS.1976.21}}

@ARTICLE{myosotis,
  author={Liu, Changxu and Zhou, Hao and Yang, Lan and Wu, Zheng and Dai, Patrick and Li, Yinlong and Wu, Shiyong and Yang, Fan},
  journal={IEEE Transactions on Computer-Aided Design of Integrated Circuits and Systems}, 
  title={Myosotis: An Efficiently Pipelined and Parameterized Multiscalar Multiplication Architecture via Data Sharing}, 
  year={2025},
  volume={44},
  number={7},
  pages={2738-2750},
  doi={10.1109/TCAD.2024.3524364}}

@software{arkworks,
  author = {arkworks contributors},
  title = {\texttt{arkworks} zkSNARK ecosystem},
  url = {https://arkworks.rs},
  year = {2022},
}

@misc{ed448_paper,
      author = {Mike Hamburg},
      title = {Ed448-Goldilocks, a new elliptic curve},
      howpublished = {Cryptology {ePrint} Archive, Paper 2015/625},
      year = {2015},
      url = {https://eprint.iacr.org/2015/625}
}

@misc{pseudo_mersenne,
      author = {Kaushik Nath and Palash Sarkar},
      title = {Efficient Arithmetic In (Pseudo-)Mersenne Prime Order Fields},
      howpublished = {Cryptology {ePrint} Archive, Paper 2018/985},
      year = {2018},
      url = {https://eprint.iacr.org/2018/985}
}

@InProceedings{kzg_pcs,
author="Kate, Aniket
and Zaverucha, Gregory M.
and Goldberg, Ian",
editor="Abe, Masayuki",
title="Constant-Size Commitments to Polynomials and Their Applications",
booktitle="Advances in Cryptology - ASIACRYPT 2010",
year="2010",
publisher="Springer Berlin Heidelberg",
address="Berlin, Heidelberg",
pages="177--194",
isbn="978-3-642-17373-8"
}

@INPROCEEDINGS{hyrax,
  author={Wahby, Riad S. and Tzialla, Ioanna and Shelat, Abhi and Thaler, Justin and Walfish, Michael},
  booktitle={2018 IEEE Symposium on Security and Privacy (SP)}, 
  title={Doubly-Efficient zkSNARKs Without Trusted Setup}, 
  year={2018},
  volume={},
  number={},
  pages={926-943},
  doi={10.1109/SP.2018.00060}}

@inproceedings{mo2023haac,
  title={Haac: A hardware-software co-design to accelerate garbled circuits},
  author={Mo, Jianqiao and Gopinath, Jayanth and Reagen, Brandon},
  booktitle={Proceedings of the 50th Annual International Symposium on Computer Architecture},
  pages={1--13},
  year={2023}
}

@inproceedings{garimella2025network,
  title={Network and Compiler Optimizations for Efficient Linear Algebra Kernels in Private Transformer Inference},
  author={Garimella, Karthik and Neda, Negar and Ebel, Austin and Jha, Nandan Kumar and Reagen, Brandon},
  booktitle={2025 IEEE/ACM International Conference On Computer Aided Design (ICCAD)},
  pages={1--10},
  year={2025},
  organization={IEEE}
}

@INPROCEEDINGS{deepraj1,
  author={Soni, Deepraj and Nabeel, Mohammed and Neda, Negar and Karri, Ramesh and Maniatakos, Michail and Reagen, Brandon},
  booktitle={2023 IEEE/ACM International Symposium on Low Power Electronics and Design (ISLPED)}, 
  title={Quantifying the Overheads of Modular Multiplication}, 
  year={2023},
  volume={},
  number={},
  pages={1-6},
  doi={10.1109/ISLPED58423.2023.10244324}}

@INPROCEEDINGS{legozk,
  author={Yang, Zhengbang and Zhao, Lutan and Li, Peinan and Liu, Han and Li, Kai and Zhao, Boyan and Meng, Dan and Hou, Rui},
  booktitle={2025 IEEE International Symposium on High Performance Computer Architecture (HPCA)}, 
  title={LegoZK: A Dynamically Reconfigurable Accelerator for Zero-Knowledge Proof}, 
  year={2025},
  volume={},
  number={},
  pages={113-126},
  doi={10.1109/HPCA61900.2025.00020}}

@inproceedings{distmsm,
author = {Ji, Zhuoran and Zhang, Zhiyuan and Xu, Jiming and Ju, Lei},
title = {Accelerating Multi-Scalar Multiplication for Efficient Zero Knowledge Proofs with Multi-GPU Systems},
year = {2024},
isbn = {9798400703867},
publisher = {Association for Computing Machinery},
address = {New York, NY, USA},
url = {https://doi.org/10.1145/3620666.3651364},
doi = {10.1145/3620666.3651364},
booktitle = {Proceedings of the 29th ACM International Conference on Architectural Support for Programming Languages and Operating Systems, Volume 3},
pages = {57–70},
numpages = {14},
location = {La Jolla, CA, USA},
series = {ASPLOS '24}
}

@inproceedings{morph,
author = {Jianming Tong and Jingtian Dang and Simon Langowski and Tianhao Huang and Asra Ali and Jeremy Kun and Srini Devadas and Tushar Krishna},
title = {MORPH: Enabling AI ASICs for Zero Knowledge Proof},
year = {2026},
booktitle = {Proceedings of the 63nd Annual ACM/IEEE Design Automation Conference},
location = {Los Angeles, California, United States},
series = {DAC '26}
}

@INPROCEEDINGS{rpu,
  author={Soni, Deepraj and Neda, Negar and Zhang, Naifeng and Reynwar, Benedict and Gamil, Homer and Heyman, Benjamin and Nabeel, Mohammed and Badawi, Ahmad Al and Polyakov, Yuriy and Canida, Kellie and Pedram, Massoud and Maniatakos, Michail and Cousins, David Bruce and Franchetti, Franz and French, Matthew and Schmidt, Andrew and Reagen, Brandon},
  booktitle={2023 IEEE International Symposium on Performance Analysis of Systems and Software (ISPASS)}, 
  title={RPU: The Ring Processing Unit}, 
  year={2023},
  volume={},
  number={},
  pages={272-282},
  doi={10.1109/ISPASS57527.2023.00034}}

@INPROCEEDINGS{ciflow,
  author={Neda, Negar and Ebel, Austin and Reynwar, Benedict and Reagen, Brandon},
  booktitle={2024 IEEE International Symposium on Performance Analysis of Systems and Software (ISPASS)}, 
  title={CiFlow: Dataflow Analysis and Optimization of Key Switching for Homomorphic Encryption}, 
  year={2024},
  volume={},
  number={},
  pages={61-72},
  doi={10.1109/ISPASS61541.2024.00016}}

@article{osiris,
author = {Ebel, Austin and Reagen, Brandon},
title = {Osiris: A Systolic Approach to Accelerating Fully Homomorphic Encryption},
year = {2026},
issue_date = {March 2026},
publisher = {Association for Computing Machinery},
address = {New York, NY, USA},
volume = {23},
number = {1},
issn = {1544-3566},
url = {https://doi.org/10.1145/3788287},
doi = {10.1145/3788287},
journal = {ACM Trans. Archit. Code Optim.},
month = mar,
articleno = {21},
numpages = {27}
}

@INPROCEEDINGS{nocap,
  author={Samardzic, Nikola and Langowski, Simon and Devadas, Srinivas and Sanchez, Daniel},
  booktitle={2024 57th IEEE/ACM International Symposium on Microarchitecture (MICRO)}, 
  title={Accelerating Zero-Knowledge Proofs Through Hardware-Algorithm Co-Design}, 
  year={2024},
  volume={},
  number={},
  pages={366-379},
  doi={10.1109/MICRO61859.2024.00035}}

\end{document}